\documentclass{appolb}
\usepackage{epsfig}

\begin{document}

\title{RADIATIVE CORRECTIONS TO THE LEFT-RIGHT MODEL\thanks{Presented at the XXIII School of Theoretical Physics, Ustro\'n'99,
Poland, September 15-22, 1999}$\;$\thanks{Work supported in part by the
Polish Committee for Scientific Research under 
Grants Nos. 2P03B08414 and 2P03B04215.}}
\author{M. CZAKON
\address {Department of Field Theory and Particle Physics, Institute 
of Physics, \\ 
University of
Silesia, Uniwersytecka 4, PL-40-007 Katowice, Poland}}
\maketitle

\begin{abstract}
The oblique part of the radiative corrections to the Left-Right model is described. The leading
non-logarithmic terms are explicitly written. It is argued, on the basis of a comparison with 
the Standard Model, that one cannot use the loop contributions of the latter to refine 
phenomenological analyses of the Left-Right model, and by the same of any general extension.
\end{abstract}

\section{Introduction}

The recent discovery of neutrino oscillations at SuperKamiokande \cite{sup} opens a new possibility of studying elementary particle models.
One of the simplest extensions of the successful Standard Model (SM), that can naturally handle massive neutrinos, through the 
see-saw mechanism, is the Left-Right Model (LRM). Although relatively simple and long standing, the LRM has only had partial studies at
the loop level. It has been known for long that higher order perturbation theory can give many surprises. The conjecture, that a model
reaching SM tree level amplitudes in the limit of large masses of the extension sector, should also have small contributions  to the SM
radiative corrections, has been recently criticized \cite{cr}.

Here we will focus on the Minimal Left-Right Model (MLRM) \cite{mlrm}, and show that despite similarities in the 
construction of the renormalization scheme, the structure of the leading oblique corrections is different. In this way we can argue that,
since even in a restricted extension of the SM, with similar features, the conjecture stated above is false, one is not allowed to
assume it in any analysis aiming at generality or completeness.

The main argument can be stated as follows. In the MLRM, the mixing angles and the gauge coupling renormalization constants can be expressed
through the electric charge and gauge boson mass counterterms. The leading non-logarithmic terms come therefore from the ratio of the
vector self-energies, just as they did in the SM, and some gauge boson masses. However there is no custodial symmetry to protect from 
 a quadratic dependence on the Higgs boson masses. Even with moderate bounds in the TeV range, they would destroy the perturbative expansion,
if the masses in the denominators were not large. With a guess, justified by a closer inspection of the origin of the different contributions,
that the scale of the denominators should be the same for all the expressions, we immediately see that the strong dependence on the top
quark mass, so cherished in the SM, will drop out. We will show that its place is taken by heavy Majorana neutrinos. The Higgs bosons will
not be dangerous any more, but they will be a major ingredient of the final result, which is just the opposite of the SM behavior.

The organization of the paper is the following. In the next section, the renormalization scheme will be introduced, and its applicability
will be discussed. The following will give a general discussion of the structure of the radiative corrections, together with a comparison
of both, the SM and the MLRM. The fourth section will give analytical results for the leading terms. Conclusions will close the main text. 
Two appendices will contain the main formulae of the gauge sector masses and mixings, indispensable for a full understanding of the argument,
and the explicit form of the fermion loop contribution to the vector boson self-energies.

\section{Renormalization}

Choosing a proper renormalization scheme \cite{heavy} is a two step process. First, we have to decide whether to renormalize the Green functions or
only scattering matrix elements. One can even take a mixed option. The main motivation is that we wish not to perform a precision test of
the LRM. Our aim is only to find an easy way of comparing its radiative corrections with those of the SM, as functions of the heavy sector masses.
To this end, it is necessary to restrict one's considerations to processes which do not need the Higgs sector to be renormalized. This follows
from the large number of parameters, that would be involved. We therefore think only of four-fermion reactions with light external states.
This allows us to forget about gauge boson wave function renormalization (apart from the photon one), a very handy simplification. All we have 
to do is to renormalize the external fermion wave functions, mixing angles and masses.

The second step is to give the renormalization conditions, which is somehow equivalent to choosing the input parameters. In the SM, the 
renormalization constants are best expressed through the gauge boson mass counterterms, meaning that we have an on-mass-shell scheme, and
through the electric charge counterterm from the definition of the fine structure constant through the Thomson scattering. Of course, since
we need high precision input parameters, the mass of the $W$ boson is fixed from the muon decay. Just the same way, the electric charge is
measured in, for example, the quantum Hall effect. Nevertheless, the possibility of expressing the renormalization constants by the masses
of the gauge bosons and the electric charge is a consequence of the fact, that in the SM, we have three parameters that decide on the structure
of the gauge sector:
\begin{equation}
g,g',v,
\end{equation}
where $v$ is the Higgs doublet vacuum expectation value and $g,g'$ are the gauge couplings. Now, in the MLRM that we consider here, the situation is
analogous. We have the following parameters (see Appendix A):
\begin{equation}
g,g',\kappa_1,\kappa_2,v_R.
\end{equation}
They correspond to the four gauge boson masses and the electric charge. The similarity of the models simplifies
the drawing of any conclusions, but misleads one to thinking that the structure of the radiative corrections should be roughly the same.

To make the discussion more concrete, we will turn our attention to the muon decay in the next section. Here we only give the necessary 
Weinberg angle counterterm:
\begin{eqnarray}
\delta s_W^2 &=& 2 c_W^2 \frac{(\delta M_{Z_2}^2+\delta M_{Z_1}^2)-(\delta M_{W_2}^2+\delta M_{W_1}^2)}
{(M_{Z_2}^2+M_{Z_1}^2)-(M_{W_2}^2+M_{W_1}^2)} \nonumber \\ & & \nonumber \\
& & +\frac{1}{2} \frac{(M_{W_2}^2+M_{W_1}^2)(\delta M_{Z_2}^2+\delta M_{Z_1}^2)
+(M_{Z_2}^2+M_{Z_1}^2)(\delta M_{W_2}^2+\delta M_{W_1}^2)}
{\left((M_{Z_2}^2+M_{Z_1}^2)-(M_{W_2}^2+M_{W_1}^2)\right)^2} \nonumber \\ & & \nonumber \\
& & -\frac{1}{2} \frac{(2 M_{Z_1}^2+M_{Z_2}^2)\delta M_{Z_1}^2+(2 M_{Z_2}^2+M_{Z_1}^2)\delta M_{Z_2}^2}
{\left((M_{Z_2}^2+M_{Z_1}^2)-(M_{W_2}^2+M_{W_1}^2)\right)^2}. 
\end{eqnarray}
The interesting feature of this expression is, as we announced in the introduction, the dependence of its denominator
on the heavy sector masses. One can check that:
\begin{equation}
\left((M_{Z_2}^2+M_{Z_1}^2)-(M_{W_2}^2+M_{W_1}^2)\right) = \frac{g^2}{2c_M^2 c_W^2} v_R^2.
\end{equation}
This comes usually as a surprise, after a cancellation in the linear expression relating the different counterterms. It only acquires a 
meaning when we think of this in view of the arguments presented at the beginning of this paper.

\section{Oblique corrections}

The presentation of the general features of the radiative corrections will be done on the example muon decay, 
since it has had many studies at different orders of perturbation theory and in 
different models. The external light neutrino states, together with the see-saw mechanism lead to the conclusion that
to a good approximation we may concentrate on the diagram with $W_1$ exchange \cite{w1}. The one-loop expressions should be 
inserted in the vertices and the gauge boson wave function and supplemented by box diagrams. The neglect of the $W_2$
diagram, has the additional advantage that we do not have to worry about infrared divergences. They have been absorbed in
the definition of $G_F$. We can factorize the correction much the same way it has been done in the SM \cite{si}:
\begin{equation}
G_F = \frac{\pi\alpha}{\sqrt{2}}\frac{1}{M_{W_1}^2s_W^2}(1+\Delta r).
\end{equation}
The quantity $\Delta r$ can be written as:
\begin{equation}
\Delta r = \frac{Re \Pi_{W_1W_1}(M_{W_1}^2)-\Pi_{W_1W_1}(0)}{M_{W_1}^2}-\Pi'_{\gamma\gamma}(0)-\frac{\delta s_W^2}{s_W^2}+\delta_{V+B}.
\end{equation}
The $\Pi$s denote the different boson self-energies and $\delta_{V+B}$ stands for the vertex and box corrections. We will not study these 
latter, although they are important. The first term on the right hand side exhibits only a roughly logarithmic dependence on the heavy particle
masses. The second has the same behavior, since it can be connected to the running of the fine structure constant. The leading terms concentrate
in the third term. The respective behavior of the first and the third term, as function of a heavy fermion mass has been depicted on Fig. 1.
It should be understood, that in any process, we will encounter only this three kind of terms. Therefore the leading terms are contained in
$\delta s_W^2$.
\begin{center}
\begin{figure}
\epsfig{figure=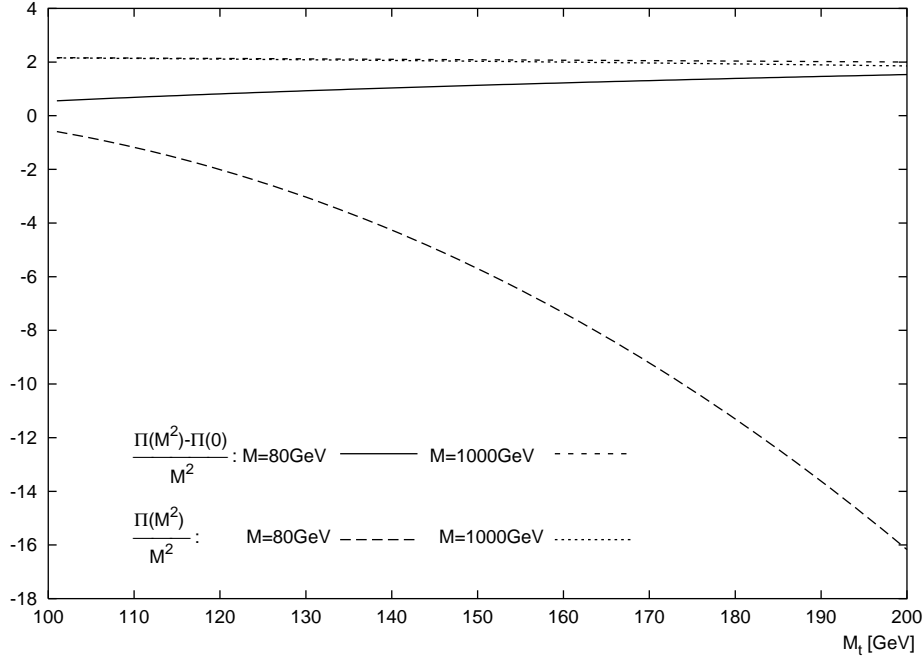}
\caption{This figure shows the behavior of the transverse part of a vector boson self energy as function of a heavy fermion mass
running in the loop. A factor of $\frac{\alpha}{4\pi}$ has been omitted, and the couplings are left-handed. Although this should
be understood as being only a qualitative picture, we note that the dependence is large solely in the case of a light boson, and
it is the value of the self-energy that matters, not its change from the zero scale up to the mass scale.}
\end{figure}
\end{center}

\section{Leading terms}

Let us present the contributions quadratic in the top mass, the heavy neutrino masses, and the Higgs boson masses as they enter $\Delta r$:
\begin{eqnarray}
(\Delta r)^{top}_{LR} &=& \frac{\sqrt{2}G_F}{8\pi^2} c_W^2\left( \frac{c_W^2}{s_W^2}-1 \right)\frac{M_{W_1}^2}
{M_{W_2}^2-M_{W_1}^2}3m_t^2, \nonumber \\
(\Delta r)^{N}_{LR} &=& \sum_{N=heavy} \frac{\sqrt{2}G_F}{16\pi^2}c_M^2 c_W^2 
\frac{M_{W_1}^2}{M_{W_2}^2-M_{W_1}^2}m_N^2, \nonumber \\
(\Delta r)^{\mbox{\it lightest Higgs}}_{LR} &=& \frac{\sqrt{2}G_F}{48\pi^2}\left(\frac{M_{W_1}^2}{M_{W_2}^2}
\frac{c_W^2}{s_W^2}(1-2s_W^2)+
\frac{M_{W_1}^2}{M_{Z_2}^2}\frac{1}{s_W^2}(4c_W^2-1)\right)M_{H^0_0}^2. \nonumber \\
\end{eqnarray}
The main features of these expressions are:
\begin{enumerate}
\item loss of quadratic dependence on the top mass, which is to mean that even with moderate bounds on the $W_2$ mass
such as $M_{W_2} > 400GeV$, the SM logarithmic terms are larger than the first contribution,
\item large contribution of heavy neutrinos, which effectively take the place the top had in the SM,
\item quadratic dependence on the heavy Higgs boson masses, which is a novelty compared to the SM, which
is protected by the custodial symmetry.
\end{enumerate}
The moral of the above can be only one. Namely, a simple extension of the SM, as is the MLRM, can be crucially different
from its tree-level high mass limit.

\section{Conclusions}

On the example of a generic, full fledged and self consistent model, we have shown, that the radiative corrections
are highly influenced by the whole structure of the theory. The differences that disappear at tree level, when the 
additional particle masses get large, do not cancel at the loop level. Worse than that, successes of the SM, like
the top mass prediction, get lost when we expand the gauge group. This should be taken as another hint of
the fact, that no general conclusion can be draw on models operating at higher mass scales, without their precise
definition and study.

\appendix

\section{}

Here we define shortly the model under consideration \cite{gu,gl}. The Higgs sector contains two triplets and a bidoublet:
\begin{equation}
\Delta_{L,R} = \left( 
  \begin{array}{cc}
  \delta^+_{L,R}/\sqrt{2} & \delta^{++}_{L,R} \\
  \delta^0_{L,R} & -\delta^+_{L,R}/\sqrt{2} \\
  \end{array} \right), \;\;\;\;
\Phi = \left(
  \begin{array}{cc}
  \phi^0_1 & \phi^+_1 \\
  \phi^-_2 & \phi^0_2 \\
  \end{array} \right).
\end{equation}
They get vacuum expectation values:
\begin{equation}
<\Delta_{L,R}> = \left( 
  \begin{array}{cc}
  0 & 0 \\
  v_{L,R}/\sqrt{2} & 0 \\
  \end{array} \right), \;\;\;\;
<\Phi> = \left(
  \begin{array}{cc}
  \kappa_1/\sqrt{2} & 0 \\
  0 & \kappa_2/\sqrt{2} \\
  \end{array} \right).
\end{equation}
We assume that $v_L$ vanishes.

The gauge sector contains two additional gauge bosons $W_2^\pm$ and $Z^0_2$. Their masses
are given by:
\begin{eqnarray}
M^2_{W_{1,2}} &=& \frac{g^2}{4}\left(\kappa_+^2+v_R^2\mp\sqrt{v_R^4+4\kappa_1^2\kappa^2_2}\right), \\
M^2_{Z_{1,2}} &=& \frac{1}{4}\left(\left((g^2\kappa_+^2+2v_R^2(g^2+g'^2)\right)\right. \\ \nonumber
              & &  \mp\left.\sqrt{(g^2\kappa_+^2+2v_R^2(g^2+g'^2))^2-4g^2(g^2+2g'^2)\kappa^2_+v_R^2}\right).
\end{eqnarray}
They are mixtures of the interaction eigenstates:
\begin{eqnarray}
\left( \begin{array}{c} W_L^\pm \\ W_R^\pm \end{array} \right) & = &
\left( \begin{array}{cc} \cos{\zeta} & \sin{\zeta} \\ -\sin{\zeta} & \cos{\zeta} \end{array} \right) 
\left( \begin{array}{c} W^\pm_1 \\ W^\pm_2 \end{array} \right),
\\ \nonumber \\ \nonumber
\left( \begin{array}{cc} W_L^3 \\ W_R^3 \\ B \end{array} \right) & = &
\left( \begin{array}{ccc} c_Wc & c_Ws & s_W \\ -s_Ws_Mc-c_Ms & -s_Ws_Ms+c_Mc & c_Ws_M \\ 
                         -s_Wc_Mc+s_Ms & -s_Wc_Ms -s_Mc & c_Wc_M \end{array} \right)
\left( \begin{array}{cc} Z_1 \\ Z_2 \\ A \end{array} \right), \\
\end{eqnarray}
where:
\begin{eqnarray}
c_W \equiv \cos{\theta_W}, \;\;\; s_W &\equiv& \sin{\theta_W}, \;\;\; c_M \equiv \frac{\sqrt{\cos{2\theta_W}}}{\cos{\theta_W}}, 
             \;\;\; s_M \equiv \tan{\theta_W} \nonumber \\
s &\equiv& \sin{\phi}, \;\;\; c \equiv \cos{\phi}.
\end{eqnarray}
These mixing angles enter all the physical observables. It is their renormalization that induces the form of the non-logarithmic 
leading terms.

\section{}

In this appendix, we give the formula for a gauge boson self-energy, coming from a fermion loop:
\begin{eqnarray}
\Pi(p^2) &=& p^2  ( 2/3 v^a_f v^b_f + 2/3 a^a_f a^b_f ) \nonumber \\
&+& B_0(p^2,(m^a_f)^2,(m^b_f)^2)  (  - 4 v^a_f v^b_f m^a_f m^b_f + 4 a^a_f a^b_f m^a_f m^b_f ) \nonumber \\
&+& B_1(p^2,(m^a_f)^2,(m^b_f)^2)p^2  ( 4 v^a_f v^b_f + 4 a^a_f a^b_f ) \nonumber \\
&+& B_{00}(p^2,(m^a_f)^2,(m^b_f)^2)  ( 8 v^a_f v^b_f + 8 a^a_f a^b_f ) \nonumber \\
&+& B_{11}(p^2,(m^a_f)^2,(m^b_f)^2)p^2  ( 4 v^a_f v^b_f + 4 a^a_f a^b_f ) \nonumber \\
&-& 2 v^a_f v^b_f (m^a_f)^2 - 2 v^a_f v^b_f (m^b_f)^2 - 2 a^a_f a^b_f (m^a_f)^2 - 2 a^a_f a^b_f (m^b_f)^2. \nonumber \\
\end{eqnarray}
The notation used, is the one of LoopTools \cite{th}. $m^a_f$ and $m^b_f$ are the fermion masses inside of the loop. $v^a_f,v^b_f,a^a_f$ and 
$a^b_f$ are their vector and axial-vector couplings. $p^2$ is the square of the external momentum.

\end{document}